\documentclass{statsoc}%
\usepackage{amsbsy,amsfonts,amscd,amsmath,amssymb}
\usepackage{bm}
\usepackage{epsfig}
\usepackage{graphicx,geometry}
\usepackage{latexsym, lscape}
\usepackage{multirow,mathrsfs}
\usepackage{natbib}
\usepackage{pifont}

\providecommand{\U}[1]{\protect\rule{.1in}{.1in}}      
\usepackage{algorithmic}
\usepackage{algorithm}

\newcommand{\LINE}{	\item[]}
 \newcommand{\sgn}{\operatorname{sgn}}

\title[Ecological State Space Models]{Ecological non-linear state space model selection via adaptive particle Markov chain Monte Carlo (AdPMCMC)}
\author[Gareth W. Peters {\it et al.}]{ Gareth W. ~Peters}
\address{UNSW Mathematics and Statistics Department, Sydney, 2052, Australia.}
\email{garethpeters@unsw.edu.au}
\author{Geoffrey R. ~Hosack}
\address{CSIRO Mathematics, Informatics and Statistics, GPO Box 1538, Hobart}
\author{Keith R. ~Hayes}
\address{CSIRO Mathematics, Informatics and Statistics, GPO Box 1538, Hobart}
{\date{Working Paper, Version from \today }}

\begin{document}
\maketitle

\begin{abstract}
We develop a novel advanced Particle Markov chain Monte Carlo algorithm that is capable of sampling from the posterior distribution of non-linear state space models for both the unobserved latent states and the unknown model parameters. We apply this novel methodology to five population growth models, including models with strong and weak Allee effects, and test if it can efficiently sample from the complex likelihood surface that is often associated with these models. Utilising real and also synthetically generated data sets we examine the extent to which observation noise and process error may frustrate efforts to choose between these models. Our novel algorithm involves an Adaptive Metropolis proposal combined with an SIR Particle MCMC algorithm (AdPMCMC). We show that the AdPMCMC algorithm samples complex, high-dimensional spaces efficiently, and is therefore superior to standard Gibbs or Metropolis Hastings algorithms that are known to converge very slowly when applied to the non-linear state space ecological models considered in this paper. Additionally, we show how the AdPMCMC algorithm can be used to recursively estimate the Bayesian Cram\'er-Rao Lower Bound of \citep{tichavský1998posterior}. We derive expressions for these Cram\'er-Rao Bounds and estimate them for the models considered. Our results demonstrate a number of important features of common population growth models, most notably their multi-modal likelihood surfaces and dependence between the static and dynamic parameters. We conclude by sampling from the posterior distribution of each of the models, and use Bayes factors to highlight how observation noise significantly diminishes our ability to select among some of the models, particularly those that are designed to reproduce an Allee effect. These result have important ramifications for ecologists searching for evidence of all forms of density dependence in population time series.

\end{abstract}


\newpage
\section{Introduction}

Ecologists in recent years have begun to fit Bayesian state space models to ecological time series,\\ \citep{Millar2000, Clark2004, Ward2006}. State space models (SSMs) avoid biases that can occur when either the observations are assumed to be perfect or the process model is treated deterministically \citep{Carrol, Shenk1998, Calder2003, Freckleton2006}. SSM inference involves jointly estimating the latent state vector and the static parameter vector of the models for observations and states. Many ecological studies that estimate density dependence using time series data, either do not include jointly estimated observation error or exclude it altogether \citep{Sibly2005, Saether2008, Gregory2010}. Studies that do incorporate process error and observation noise into ecological inference sometimes use very special cases of linear state and observation equations \citep{Lindley, Williams2003, Linden2009}. More realistic, non-linear models may be implemented sub-optimally in an Extended Kalman Filter \citep{Wang2007latent, Zeng, welch1995introduction}. Linear approximations of non-linear state equations \citep{Wang2007latent} can allow for a combination of suboptimal Kalman filtering and Gibbs sampling to be performed. Linear approximation, however, can lead to inaccurate parameter posterior inference for non-linear SSM with non-negligible observation noise and process error.

Studies that employ the Metropolis-Hastings within Gibbs (MHG) algorithm avoid these restrictive linear assumptions or local linearizations, \citep{Millar2000, Ward2006, Clark2004, Saether2007}, but the performance of MHG is known to be very sensitive to the process model and observation model parameterisation as discussed in \citep{AndrieuDoucetHolenstein2010}. The standard MHG algorithm can be difficult to tune for non-linear ecological SSMs, \citep{Millar2000}, for two reasons. Firstly, non-negligible posterior correlations can occur in the high dimensional parameter space that is composed of both the latent state process and the static parameters. Secondly, the joint likelihood surfaces of the static parameters for popular non-linear population growth models (irrespective of observation error) can be multi-modal and have strong sharp flat ridges, \citep{Polansky2009}. Furthermore, an MHG algorithm that has not sufficiently mixed over modes and ridges will corrupt all model selection and model averaging routines.

In this paper, we describe a novel technical development that is specifically designed to address non-linear dynamics in SSMs. The method uses a Particle MCMC algorithm for joint process and parameter estimation in non-linear and non-Gaussian SSMs \citep{AndrieuDoucetHolenstein2010}, coupled to an adaptive Metropolis proposal which we denote by the Adaptive Particle Markov chain Monte Carlo algorithm (AdPMCMC). We apply this novel methodology to synthetic and real data sets. Our objectives are to examine the performance of the new algorithm in non-linear population growth models, with complex likelihood surfaces, and to re-examine the problems of model selection with Bayes factors estimated via Markov chain samples. We find that the novel methodology is capable of efficiently sampling highly non-linear models without careful tuning in very high dimensional posterior models. 

Complicated likelihood surfaces with multimodal characteristics present a serious challenge to Bayesian posterior samplers, \citep{Polansky2009}. The failure of the basic MHG sampler to mix over the support of the posterior in such situations is a known concern in this case for ecological modellers and in general a well known problem in the statistics literature, see discussions and illustrations in \citep{carlin1997bayes, robert2001bayesian, bishop2006pattern}. Our AdPMCMC methodology will be shown to be capable of efficiently exploring the posterior distributions obtained from typical ecological growth models, overcoming many of the poor mixing properties of the basic MHG sampler in this context. Additionally, we use Bayes factors to highlight the ambiguity in model structure when fitting models to short ecological time series and therefore affirm that predictions outside the observed data made by a single best fitting model can be misleading. This is especially important when the intention is to capture effects that are difficult to observe, such as the Allee effect. It is precisely in estimation of quantities such as Bayes factors, that one requires a sampling methodology capable of efficiently delivering samples from the entire support of the posterior distribution.
\vspace{-0.5cm}
\subsection{Contribution}
This paper addresses three main concerns about model fitting to ecological time series. The first of these is the inclusion of observation error. \citet{Wang2007latent}, argues that observation error should be included in ecological modelling when estimating abundance of animals because measurement errors arise from several sources, for example unaccounted effects of weather on the probability of detecting animals or random failures of trapping devices. Ignoring observation error can lead to biased estimates of the parameters \citep{Carrol}. We argue it will also have an important effect on model selection and the ability to detect ecological effects such as density dependence and the strength of an Allee effect. 

The second issue involves model simplification. It is difficult to jointly estimate the latent process with the model parameters, especially when the state or observation process are non-linear and there is strong dependence between model parameters and the latent process. The approach we present will allow us to work directly with realistic non-linear models in the presence of arbitrarily high process and observation noise settings. Moreover, we do not introduce any approximation error as we have no need to linearise such models to perform filtering, such as in the approaches discussed in \citep{Wang2007latent}. 

The third issue relates to efficient estimation and robust model selection. To date relatively simple Metropolis-Hastings and Gibbs sampling routines have been used to fit ecological state space models \citep{Millar2000, Clark2004, Ward2006, Saether2007}. These algorithms are prone to convergence difficulties for two reasons: the first is due to non-negligble posterior correlations that will occur in the high dimensional parameter space that is composed of both the latent state process and the static parameters; the second is due to posterior multimodality. The AdPMCMC we propose is specifically designed for joint process and parameter estimation in non-linear and non-Gaussian states space models.
\vspace{-0.5cm}
\subsection{Structure and Notation}
The paper is structured as follows. In Section 2, we present five common population growth models, including four that are non-linear in the states, parameters or both. Two of these models include weak or strong Allee effects which can be important in ecological populations \citep{Dennis2002}. In Section 3, we summarise a number of issues regarding Bayesian estimation, prior selection and the likelihood surface that are important and relevant to the issues identified above. Section 4 presents the details of the estimation and model selection procedure, including the AdPMCMC algorithm. Section 5 presents the results and discussion which is split into two parts. The first part considers results for synthetic data and the performance of the algorithm. The second part presents the results for two real data sets under the newly developed sampling methodology. The paper concludes with brief discussion and recommendations for future developments including more sophisticated adaption routines and non-Gaussian error structures.

The notation used throughout this paper will involve: capitals to denote random variables and lower case to denote realizations of random variables; bold face to denote vectors and non-bold for scalars; sub-script will denote discrete time, where $n_{1:T}$ denotes $n_1,\ldots,n_T$. In the sampling methodology combining MCMC and particle filtering (Sequential Monte Carlo - SMC), we use the notation $[\cdot](j,i)$ to denote the j-th state of the Markov chain for the i-th particle in the particle filter. Note, the index $i$ is dropped for quantities not involving the particle filter. We will define a proposed new state prior to acceptance at iteration $j$ of the Markov chain by $[\cdot](j)'$. In addition, we note that $\bm{\theta}$ generically represents the particular model parameters under consideration, for example $M_0$ has $\bm{\theta}=\left(b_0,\sigma_{\epsilon},\sigma_{w}\right)$ and $M_4$ will have $\bm{\theta}=\left(b_5, b_6, b_7,,\sigma_{\epsilon},\sigma_{w}\right)$.

\section{Ecological State Space Models}
This section presents the ecological state space models considered, with discussion from the ecological modelling perspective, regarding density dependence and weak or strong Allee effects. 
\subsection{Observation equations}
The generic observation equation we consider is,
\begin{equation}
        y_{t} = g_{t}\left(n_{t}\right) + w_{t}
\end{equation}
where $w_{t} \sim N\left( 0,\sigma _{w}\right)$.
The observation model acknowledges that we typically do not observe the entire population of interest, but rather a sample realization and that the observation mechanisms we use are imperfect. It therefore reflects all sources of variability introduced by the data-generating mechanism. The Gaussian observation error assumption for log transformed abundances is common and widely applied in ecological contexts, e.g. \citep{Clark2004}. Importantly, this assumption can be easily extended to include more flexible classes of observation noise, such as $\alpha$- stable models \citep{PetersFanSisson2009} or others recently proposed in the ecological literature \citep{Stenseth2003, Ward2006}, under the methodological framework that we develop in this paper.

\subsection{Process equations}
\label{stateEqn}
We consider five different stochastic population dynamic models that encapsulate different ecological effects such as density dependendent mortality and strong and weak Allee effects. Each of these process models describes how the number of individuals in a population affect the subsequent growth of the population. In these models $N_t$ represents a continous latent random variable for the population size. The process error in these models ($\epsilon_{t} \sim N(0,\sigma_{\epsilon}^2)$) reflects all sources of variability in the underlying population growth process that is not captured by the model. Process error in all of the models is assumed to behave multiplicatively on the natural scale, see \citep{Clark2007}. 

\subsubsection{The exponential growth equation ($M_0$)}
The exponential growth model provides a simple density independent model for comparison with potentially more realistic density dependent models. The discrete, exponential growth log transformed equation is
	\begin{equation}
		\log N_{t + 1} = \log N_t + b_0 +\epsilon_{t}.
	\end{equation}
where $b_{0} = r$ is the maximum per-individual growth rate (defined as the difference between the per individual birth and death rates). 

\subsubsection{The Ricker equation ($M_1$)}
The discrete Ricker equation is a density dependent model that is similar to the familiar logistic growth model. After log transformation it is given by,
	\begin{equation}
		\log{N_{t + 1}} = \log{N_t} + b_0 + b_{1}N_{t} + \epsilon_{t},
	\end{equation}
where $b_{0}$ is the density-independent growth rate and $b_{1}$ governs the strength of density dependence. For populations that exhibit negative density dependence ($b_{1} < 0$), the ``carrying capacity'' of the environment (usually denoted $K$) is defined by the stable equilibrum at $-\frac{b_0}{b_1}$ so long as the density independent growth rate is positive ($b_{0} > 0$). If $b_0 < 0$ then the only stable equilibrium is located at $0$. Whilst linear in its parameters, this model is non-linear in the latent state because it contains an exponential term in $N_t$.

\subsubsection{The theta-logistic equation ($M_2$)}
\citep{Lande2003} recommend the theta-logistic equation when the form of the density dependence in the population dynamics is unknown, given here on the log scale
	\begin{equation}
		\log{N_{t + 1}} = \log{N_t} + b_0 + b_{2}\left(N_{t}\right)^{b_3} + \epsilon_{t},
	\end{equation}
where $b_3$ determines the form of density dependence. For the carrying capacity ($K = \left(-\frac{b_0}{b_2}\right)^{\frac{1}{b_3}}$) to exist, $b_0$ and $b_2$ must be of opposing sign, and it is stable only if $b_0$ and $b_3$ are of the same sign.

\subsubsection{The ``mate-limited'' logistic equation ($M_3$)}
\citep{Morris2002} propose a ``mate-limited'' Allee effect equation, presented here under a log transformation,
	\begin{equation}
		\log{N_{t + 1}} = 2\log{N_t} - \log\left(b_{4} + N_t\right) + b_0 + b_{1}N_t + \epsilon_{t},
	\end{equation}
where $b_4 > 0$ represents the population size at which the per-individual birth rate is half of what it would be if mating was not limited, thereby controling the population size at which Allee effects are ``noticeable''. \citep{Dennis2002} notes that mate-limited models can describe, in a phenomenological fashion, Allee effects from other biological mechanisms besides mate limitation. 

\subsubsection{The ``Flexible-Allee'' logistic equation ($M_4$)}
This modified version of the Ricker model allows for both strong and weak Allee effects,
	\begin{equation}
		\log{N_{t + 1}} = \log{N_t} + b_{5} + b_{6}N_{t} + b_{7}N_{t}^2 + \epsilon_{t},
	\end{equation}
where $K = \frac{\left(-b_{6} - \sqrt{b_{6}^2 - 4b_{5}b_{7}}\right)}{2b_{7}}$, and $C = \frac{\left(-b_{6} + \sqrt{b_{6}^2 - 4b_{5}b_{7}}\right)}{2b_{7}}$. If these roots are real, then $C$ represents the threshold of population size below which per capita population growth is negative. In the deterministic model, when $0 < C < K$, the Allee effect is strong and it is possible for an unstable equilibrium at $N = C$ to occur between two stable equilibria, $N = 0$ and $N = K$. When $C < 0$, the Allee effect is weak and a single stable equilibria occurs at $N = K$. If $C$ and $K$ are not real, then $N=0$ is the only stable equilibrium. This model is a discrete version of a continuous time model previously used in theoretical studies of spatial population dynamics \citep{Lewis1993}; see \citep{Boukal2002} for further discussion of Allee effects in discrete time population models. Note that model $M_4$ differs from model $M_3$ in that the latter can only exhibit a strong Allee effect.

The process models $M_0$ to $M_4$ are nested models. Specifically, setting $b_7 = 0$ in model $M_4$ and $b_4 = 0$ in model $M_3$ returns model $M_1$. Setting $b_3 = 1$ in model $M_2$ also returns model $M_1$. Setting $b_{1} = 0$ in model $M_1$ returns the density independent model $M_0$.

\section{Bayesian Estimation}
Our approach estimates the latent process and model parameters for models $M_0$ through to $M_4$ using advanced Bayesian inference. The Markov stochastic process $N_{0:T}$ is unobserved (latent) and must be estimated over the time interval $\left[0,T\right]$ at discrete time points $t=0$ through to $T$. In this setting we are interested in estimating the entire data set as opposed to a sequential, real-time estimates of the latent states jointly with the model parameters. As such the posterior of interest to this problem is given by $p(\bm{\theta},n_{1:T}|y_{1:T},M_i)$, where we have absorbed the initial state value ($N_0$) into the vector of static parameters $\bm{\theta}$.

\subsection{Priors, Likelihood and Posterior} \label{Priors}

Table \ref{tab1} summarises the priors selected for the process and observation model parameters. The process model parameters $b_0, b_1, b_2, b_3, b_5, b_6$ and $b_7$ are given Gaussian distribution priors. The prior for $b_4$ is assumed to be nonnegative \citep{Morris2002} and is given a Gamma prior distribution. In addition, we give the noise variances $\sigma _{\epsilon}^{2}, \sigma _{w}^{2}$ inverse gamma priors with the parameters $\alpha_{\epsilon}= \alpha_{w} = \frac{T}{2}$ and $\beta_{\epsilon}= \beta_{w} = \frac{2\left(\alpha_{\epsilon}-1\right)}{10}$.

\begin{table}
\caption{\label{tab1} Priors for the parameters.}
\centering
\fbox{
\begin{tabular}{*{2}{c}}
Parameters & Priors \\ \hline
$b_{0, 1, 2, 3, 5, 6, 7}$ & $N\left(0, 1\right)$ \\
$b_{4}$ & $G\left(shape = 1, scale = 10\right)$ \\
$\sigma _{\epsilon}^{2}$ & $IG\left(\alpha_{\epsilon},\beta_{\epsilon}\right)$ \\
$\sigma _{w}^{2}$ & $IG\left(\alpha_{w},\beta_{w}\right)$ \\
\end{tabular}}
\end{table}

We also assume that observations are independent and identically distributed. Hence, the generic likelihood is given by
\begin{equation}
\mathcal{L}\left(n_{1:T},\bm{\theta};y_{1:T}\right) = \prod_{t=1}^{T}p\left(y_{t}|n_{t},\sigma_{w}^{2}\right),
\end{equation}
where $p\left( y_{t}|n_{t},\sigma_{w}^{2}\right)$ is a Gaussian distribution with mean $n_{t}$ and unknown variance $\sigma_{w}^2$. We assume that the observations are conditionally independent given the latent state, with latent states under a first order Markov dependence. This produces a target posterior distribution given by
\begin{equation}
\begin{split}
p\left( \bm{\theta},n_{1:T}|y_{1:T}\right) &\propto
p\left(\bm{\theta }\right) \prod^T_{t=1} p\left(y_t|n_t, \bm{\theta}\right)p\left(n_t|n_{t-1}, \bm{\theta}\right).
\end{split}
\label{PosteriorStateParam}
\end{equation}

\subsection{Estimation and Model Selection}
Having specified the posterior distribution, inference and model selection proceeds by estimating popular Bayesian statistics such as the Minimum Mean Square Error (MMSE = posterior mean) and evaluating the posterior evidence for each model. In our context this requires estimates of the following quantities
\begin{equation}
\begin{split}
\text{MMSE:  } \left(\widehat{N}^{MMSE}_{1:T},\widehat{\bm{\Theta}}^{MMSE}\right) &= \int \bm{\Theta} \int N_{1:T} \, p(\bm{\theta},n_{1:T}|y_{1:T},M_i) \,dN_{1:T}\, d\bm{\Theta},\\
\text{Posterior evidence $M_i$:   } p(y_{1:T}|M_i) &= \int p(y_{1:T}|\bm{\theta},n_{1:T},M_i)\, p(\bm{\theta},n_{1:T}|M_i) \,dN_{1:T}\, d\bm{\theta}.\\
\end{split}
\label{BayesianEstimators}
\end{equation}
Estimating these quantities for non-linear state space models presents a considerable statistical challenge. The dimension of the posterior distribution is dim$\left(\bm{\Theta}\right)$ + T, hence for moderate T $\left(\approx 50 \text{ to } 100\right)$ the dimension is large. Controlling the variance of these Bayesian estimators requires efficient posterior sampling methods from the posterior distribution, which we achieve via our adaptive version of the PMCMC algorithm of \citep{AndrieuDoucetHolenstein2010}. 

\subsection{Adaptive MCMC within Particle MCMC (AdPMCMC).}
The aim of this section is to present a novel methodology to sample from the posterior distribution given in Equation (\ref{PosteriorStateParam}), based on a version of a recent sampler specifically developed for use in state space models; see \citep{AndrieuDoucetHolenstein2010}. The methodology is known as PMCMC, and it represents a state of the art sampling framework for state space problems. These samples can then be used to form Monte Carlo estimates for Equations (\ref{BayesianEstimators}). The fundamental innovation we present in this paper is to combine an Adaptive MCMC algorithm within the PMCMC framework.

The key advantage of the PMCMC algorithm is that it allows one to jointly update the entire set of posterior parameters $\left(\bm{\Theta},N_{1:T}\right)$ and only requires calculation of the marginal acceptance probability in the Metropolis-Hastings algorithm. PMCMC achieves this by embedding a particle filter estimate of the optimal proposal distribution for the latent process into the MCMC algorithm. This allows the Markov chain to mix efficiently in the high dimensional posterior parameter space because the particle filter approximation of the optimal proposal distribution in the MCMC algorithm, thereby allowing high-dimensional parameter block updates even in the presence of strong posterior parameter dependence. The models considered in this paper for example have between three and seven static parameters to be estimated jointly with the latent state space, resulting in an additional $T$ parameters, where $T$ can range from the tens to the hundreds, or even thousands, depending on the data set considered.

In the state space setting, the Particle MCMC algorithm used to sample from the target distribution $p\left(\bm{\theta},N_{1:T}|y_{1:T}\right)$ proceeds by mimicking the marginal Metropolis-Hastings algorithm in which the acceptance probability is given by,
\begin{align*}
\alpha\left([\bm{\theta},n_{1:T}](j-1),[\bm{\theta},n_{1:T}](j)'\right)= \min \left( 1,
\frac{p\left([\bm{\theta}](j)'|y_{1:T}\right)q\left([\bm{\theta}](j)',[\bm{\theta}](j-1)\right)}{p\left([\bm{\theta}](j-1)|y_{1:T}\right)q\left([\bm{\theta}](j-1),[\bm{\theta}](j)' \right)} \right).
\end{align*}
where $q\left([\bm{\theta}](j-1),[\bm{\theta}](j)\right)$ is the proposal distribution of the PMCMC generated Markov chain for the static parameters to propose a move from state $[\bm{\theta}](j-1)$ at iteration $j-1$ to a new state $[\bm{\theta}](j)$ at iteration $j$.

To achieve this we split the standard Metropolis Hastings proposal distribution into two components. The first constructs a proposal kernel via an adaptive Metropolis scheme (\citep{roberts2009examples}, \citep{atchadé2005adaptive}) that is used to sample the static parameters $\bm{\Theta}$. Introducing an adaptive MCMC proposal kernel into the Particle MCMC setting allows the Markov chain proposal distribution to adaptively learn the regions of the marginal posterior distribution of the static model parameters that have the most mass. This significantly improves the acceptance probability of the proposal distribution and enables much more rapid and efficient mixing of the Markov chain for a small number of particles L and a simple particle filter algorithm. The second component of the proposal kernel constructs an estimate of the posterior distribution of the latent states, $N_{1:T}$, allowing us to sample a proposed trajectory. This proposal kernel is constructed via a Sequential Monte Carlo algorithm that is based on a simple SIR filter, (\citep{gordon1993novel}, \citep{doucet2009tutorial}). Note the SIR filter is suitable for PMCMC applications since it is only used as a proposal distribution and not as an empirical estimate of the posterior as is more common; see discussion of standard application in \citep{doucet2009tutorial}. In particular the proposal kernel approximates the optimal choice
\begin{equation}
\begin{split}
q\left([\bm{\theta},n_{1:T}](j-1),[\bm{\theta},n_{1:T}](j)'\right)
=q\left([\bm{\theta}](j-1),[\bm{\theta}](j)'\right)p\left([n_{1:T}](j)'|y_{1:T},[\bm{\theta}](j)'\right),
\end{split}
\label{proposalOpt}
\end{equation}
via an adaptive MCMC proposal for $q\left([\bm{\theta}](j-1),[\bm{\theta}](j)'\right)$ and a particle filter (SMC) estimate for \\ $p\left([n_{1:T}](j)'|y_{1:T},[\bm{\theta}](j)'\right)$. The SMC algorithm proposal samples (approximately) from the sequence of distributions, $\{p(n_{1:t}|y_{1:t},\bm{\theta})\}_{t=1:T}$. For a recent review of SMC methodology, of which there are several different methodologies, see \citep{doucet2009tutorial}. 

When this proposal is substituted into the standard Metropolis Hastings acceptance probability, several terms cancel to produce,
\begin{align*}
\alpha\left([\bm{\theta},n_{1:T}](j-1),[\bm{\theta},n_{1:T}](j)'\right) &= 
\min \left( 1,\frac{p\left(y_{1:T}|[\bm{\theta}](j)'\right)p\left([\bm{\theta}](j)'\right)q\left([\bm{\theta}](j)',[\bm{\theta}](j-1)\right)}{p\left(y_{1:T}|[\bm{\theta}](j-1)\right)p\left([\bm{\theta}](j-1)\right)q\left([\bm{\theta}](j-1),[\bm{\theta}](j)'\right)}
\right).
\end{align*}
We can now detail a generic version of the AdPMCMC methodology we developed.

\subsubsection{Generic Particle MCMC}
One iteration of the generic AdPMCMC algorithm proceeds as follows:

\begin{enumerate}
\item{Sample $[\bm{\theta}](j)' \sim q\left([\bm{\theta}](j-1),\cdot\right)$ from an Adaptive MCMC proposal.(Appendix 1. Algorithm 2).}
\item{Run an SMC algorithm with $L$ particles to obtain:
\begin{equation}
\begin{split}
&\widehat{p}\left(n_{1:T}|y_{1:T},[\bm{\theta}](j)'\right)
=\sum_{i=1}^{L}W_{1:T}^{(i)}\delta_{[n_{1:T}](j,i)'}\left(n_{1:T}\right)\\
&\widehat{p}\left(y_{1:T}|[\bm{\theta}](j)'\right) = \prod_{t=1}^T
\left(\frac{1}{L}\sum_{i=1}^{L} w_t\left([n_{t}](j,i)'\right)
\right)
\end{split}
\label{margLHEst}
\end{equation}
Then sample a candidate path $[N_{1:T}](j)' \sim
\widehat{p}\left(n_{1:T}|y_{1:T},[\bm{\theta}](j)'\right)$. (see Appendix 1. Algorithm 1)}
\item{Accept the proposed new Markov chain state comprised of
$[\bm{\theta},N_{1:T}](j)'$ with acceptance probability
given by
{\small{
\begin{align}
&\alpha\left([\bm{\theta},n_{1:T}](j-1),[\bm{\theta},n_{1:T}](j)'\right)
 = \min \left( 1, \frac{\widehat{p}\left(y_{1:T}|[\bm{\theta}](j)'\right)p\left([\bm{\theta}](j)'\right)q\left([\bm{\theta}](j)',[\bm{\theta}](j-1)\right)} {\widehat{p}\left(y_{1:T}|[\bm{\theta}](j-1)\right)p\left([\bm{\theta}](j-1)\right)q\left([\bm{\theta}](j-1),[\bm{\theta}](j)'\right)} \right)
\end{align}
}}
where $\widehat{p}\left(y_{1:T}|[\bm{\theta}](j-1)\right)$ is obtained
from the previous iteration of the PMCMC algorithm.}
\end{enumerate}

The key advantage of this approach is that the difficult problem of designing high dimensional proposals has been replaced with the simpler problem of designing low dimensional mutation kernels in the Sequential Monte Carlo algorithm embedded in the MCMC algorithm. This sampling approach in which Sequential Monte Carlo is used to approximate the
marginal likelihood in the acceptance probability has been shown to have several theoretical convergence properties. In particular the empirical law of the particles converges to the true filtering distribution at each iteration as a bounded linear function of time $t$ and the number of particles $L$; see \citep{AndrieuDoucetHolenstein2010}. This means is is possible to construct and efficiently sample approximately optimal path space proposals with linear cost.

\subsubsection{Generic Adaptive MCMC}
We utilise the adaptive MCMC algorithm to learn the proposal distribution for the static parameters in our posterior $\bm{\Theta}$. There are several classes of adaptive MCMC algorithms; see \citep{andrieu2006efficiency}. The distinguishing feature of adaptive MCMC algorithms, (compared to standard MCMC), is that the Markov chain is generated via a sequence of transition kernels. Adaptive algorithms utilise a combination of time or state inhomogeneous proposal kernels. Each proposal in the sequence is allowed to depend on the past history of the Markov chain generated, resulting in many possible variants.

When using inhomogeneous Markov kernels it is particularly important to ensure the generated Markov chain is ergodic, with the appropriate stationary distribution. Several recent papers proposing theoretical conditions that must be satisfied to ensure ergodicity of adaptive algorithms include, \\ \citep{andrieu2006ergodicity} and \citep{haario2005componentwise}. In particular \citep{roberts2009examples} proved ergodicity of adaptive MCMC under conditions known as \textit{Diminishing Adaptation}
and \textit{Bounded Convergence}. 

It is non-trivial to develop adaption schemes that are easily verified to satisfy these two conditions. In this paper use a mixture proposal kernel known to satisfy these two ergodicity conditions when unbounded state spaces and general classes of target posterior distributions are utilised; see \citep{roberts2009examples}. 

\subsection{Adaptive Metropolis within SIR Particle MCMC (AdPMCMC).}
In this section (and Appendix 1) we present the details of the Adaptive
Metropolis within Particle MCMC (AdPMCMC) algorithm used to sample from the
posterior on the path space of our latent states and model parameters. This involves specifying the details of the proposal distribution in Equation \ref{proposalOpt}. The proposal, $q\left([\bm{\theta}](j-1),[\bm{\theta}](j)'\right)$, involves an adaptive Metropolis
proposal comprised of a mixture of Gaussians, one component of
which has a covariance structure that is adaptively learnt
on-line as the algorithm progressively explores the posterior
distribution. The mixture proposal distribution for parameters
$\bm{\theta}$ is given at iteration $j$ of the Markov chain by,
\begin{equation}
q_j\left([\bm{\theta}](j-1),\cdot\right)= w_1
N\left(\bm{\theta};[\bm{\theta}](j-1),\frac{\left(2.38\right)^2}{d}\Sigma_j\right)
+ \left(1-w_1\right)
N\left(\bm{\theta};[\bm{\theta}](j-1),\frac{\left(0.1\right)^2}{d}I_{d,d}\right).
\label{AdaptiveProp}
\end{equation}
Here, $\Sigma_j$ is the current empirical estimate of the
covariance between the parameters of $\bm{\theta}$ estimated using
samples from the Particle Markov chain up to time $j$. The
theoretical motivation for the choices of scale factors 2.38, 0.1
and dimension d are all provided in \citep{roberts2009examples}
and are based on optimality conditions presented in \citep{roberts2001optimal}.

The proposal kernel for the latent states $N_{1:T}$, given by
$p\left([n_{1:T}](j)'|y_{1:T},[\bm{\theta}](j)'\right)$, uses the simplest
SMC algorithm known as the SIR algorithm. Therefore the mutation kernel, is given by the process model, in our case $N\left(N_t; f_{t}\left(N_{t-1},\bm{\theta}\right),
\sigma_{\epsilon}^2\right)$.

\section{Results and Analysis}
This section is split into four subsections, the first two subsections study the performance and estimation accuracy of the AdPMCMC algorithm using synthetic data generated from models $(M_0,\ldots,M_4)$ with known parameters.  We gauge the accuracy of the algorithm by comparing the marginalised average estimated mean square error (MSE) of the MMSE estimate of the latent state process $N_{1:T}$ (estimated over 20 blocks per data set with 20 data sets for the average) to a recursively defined Bayesian Cramer-Rao Lower Bound (BCRLB). In the third subsection, we study the performance of Bayes factor estimates for each of the models using evidence estimates derived from Markov chain samples obtained via the AdPMCMC algorithm. In the final sub-section, we perform parameter estimation and model selection using two real data sets.

Each iteration of the AdPMCMC algorithm an approximation to the optimal proposal is constructed and sampled to produce a proposed Markov chain state update in dimension $\text{dim}\left(\bm{\Theta}\right) + T$. Additionally, all simulation studies in this paper involved generation of a Markov chain via the AdPMCMC algorithm with the following three stages:
\begin{itemize}
\item[\textbf{stage 1:}] ANNEALED PMCMC - This stage involves a random initialization of the posterior parameters $\left(\bm{\theta},N_{1:T}\right)$ followed by an annealed AdPMCMC algorithm, using a sequence of distributions $$p_n(\bm{\theta},N_{1:T}|y_{1:T}) = p(\bm{\theta},N_{1:T})^{1-\gamma_n}p(y_{1:T}|\bm{\theta},N_{1:T})^{\gamma_n},$$ where $\gamma_n$ increases linearly from $10^{-5}$ to $1$ for the first 5,000 AdPMCMC steps. The annealing must also be integrated into the construction of the particle filter proprosal, and impacts directly on the estimate of the marginal likelihood for $\bm{\theta}$.;
\item[\textbf{stage 2:}] NON ADAPTIVE PMCMC - This stage involves a burn-in chain of 5,000 iterations in which a non-adaptive proposal is utilised for the static parameters $\bm{\theta}$ based on the non-adaptive mixture component in proposal Equation (\ref{AdaptiveProp}) and the SIR particle filter for the latent process states. These burn-in samples are used to form an initial estimate of the covariance matrix for the first iteration of the adaptive MCMC proposal;
\item[\textbf{stage 3:}] ADAPTIVE AdPMCMC - This stage involves generating samples using the SIR particle filter proposal and the mixture adaptive Metropolis proposal Equation (\ref{AdaptiveProp}) for 50,000 samples that are subsequently used in estimating model parameters. 
\end{itemize}

In addition to these burnin stages one could utilise a combination of tempering throughout the PMCMC chain. However, we note that when performing the tempering stage on the marginal likelihood, one should be careful to avoid bias in the acceptance probability. One way to overcome this problem and to still maintain all the Markov chain samples would be to perform post processing of the tempered samples, with temperature not equal to one, under the Importance Tempering framework of \citep{gramacy2010importance} within a tempered PMCMC algorithm. It should also be noted, that if performing this additional tempering, one must be very careful about the inclusion of adaptive MCMC for the static parameters and the conditions of bounded convergence and diminishing adaptation should be verified. Finally, we point out that adaption of the mutation kernel in the SMC component of the algorithm can be performed arbitrarily without any constraint on the adaption rate and one possible approach to consider involves the methodology of \citep{cornebise2008adaptive}.

\subsection{Simulation Study - Synthetic Data $M_2$ - Theta Logistic.}

Here we study the performance of the AdPMCMC algorithm in the context of a challenging non-linear process model: the theta-logistic ($M_2$). We randomly generate data sets from the model, and we show that the AdPMCMC sampling methodology produces a Markov chain that mixes efficiently and can accurately obtain estimates of Equation (\ref{BayesianEstimators}), even in the presence of significant observation noise.

\begin{figure}[!ht]
\centerline{\includegraphics[width=0.8\textwidth, height=5cm]{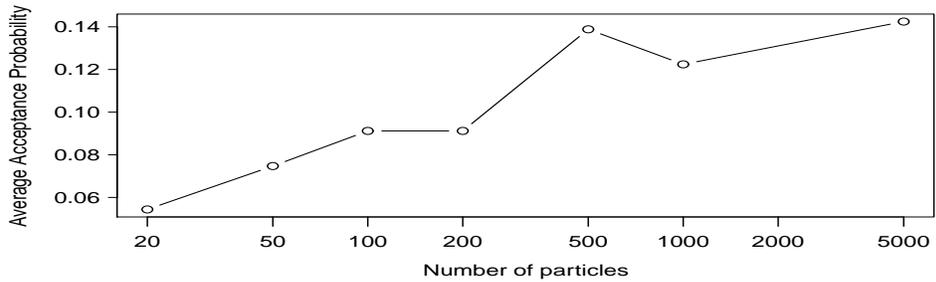}}
\caption{\footnotesize{Average acceptance probabilities of the AdPMCMC algorithm proposing updates of the MCMC in 56 dimensional space at each iteration, averaged over 50,000 Markov chain samples post burn-in, as a function of the number of particles $L$.}}
\label{fig1:AvgAccProb}
\end{figure}

\begin{figure}[!ht]
\centerline{\includegraphics[width=0.8\textwidth, height=8cm]{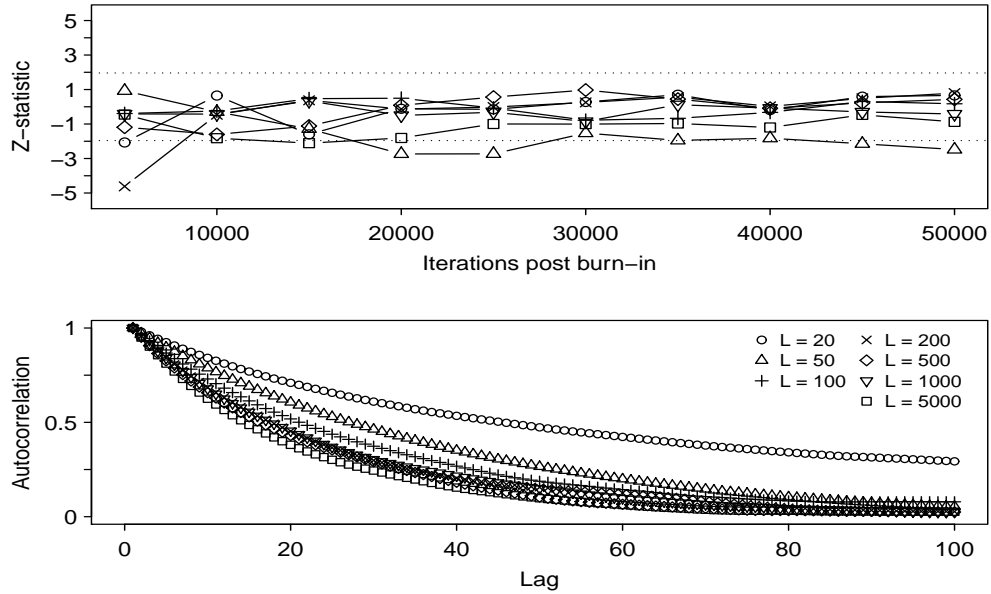}}
\caption{\footnotesize{Diagnostics for the AdPMCMC algorithm calculated post burn-in over 50,000 Markov chain samples. TOP SUBPLOT: Geweke Z-score statistics for noise $\sigma_{w}$ versus $L$. BOTTOM SUBPLOT: ACF function of the noise $\sigma_{w}$ versus the number of particles $L$. }}
\label{fig2:UniformativeThetaACF}
\end{figure}

The parameters used for the synthetic studies in this first subsection are $(b_0 = 0.15, b_2 = -0.125, b_3 = 0.1, (K = 6.2, \theta = 0.1), \sigma_{w} = 0.39, \sigma_{\epsilon} = 0.47, n_0 = 1.27)$. We produce $T=50$ observations using the simulated values of the latent process. We begin the analysis of the AdPMCMC algorithm for a range of particle numbers $L\in\left\{20,50,100,200,500,1000,5000\right\}$. Figure \ref{fig1:AvgAccProb} shows the average acceptance probability of the AdPMCMC algorithm, which proposes updates of the entire Markov chain state $\left(N_{1:T},\bm{\theta}\right)$ in 56 dimensions, at each iteration of the AdPMCMC sampler, as a function of L. As expected, increasing the number of particles results in an improved estimate of the optimal proposal distribution $\widehat{p}\left(n_{1:T}|y_{1:T},\bm{\theta}'\right)$ in Equation (\ref{proposalOpt}) and a lower variance estimate for marginal likelihood $\widehat{p}\left(y_{1:T}|\bm{\theta}'\right)$ in Equation (\ref{margLHEst}), ultimately improving the acceptance probability. The results demonstrate that it is reasonable to perform simulations with $L=500$ particles which produces efficient mixing in the AdPMCMC sampler. \\ 
\vspace{-1cm}
The bottom panel of Figure \ref{fig2:UniformativeThetaACF} presents the autocorrelation function (ACF) for $\sigma_{w}^2$ for different numbers of particles $L$. The top panel plots the Geweke Z-score time series diagnostic \citep{geweke1992evaluating} for the $\sigma_{\epsilon}^2$ parameter as a function of $L$. The settings for this diagnostic considered the difference between the means of the first 10\% and the last 50\%. This is calculated for the Markov chain at lengths of $t\in\left\{5k, 10k, \ldots, 50k\right\}$. We note that technically this diagnostic is derived for a non-adaptive Markov chain proposal. We apply this result here as a guide to convergence arguing it still produces informative results once the rate of adaption slows down, because the covariance structure used in the proposal becomes close to constant. This occurs once the chain has mixed sufficiently over the posterior support. 

\begin{figure}[!ht]
\centerline{\includegraphics[width=0.8\textwidth, height=7.5cm]{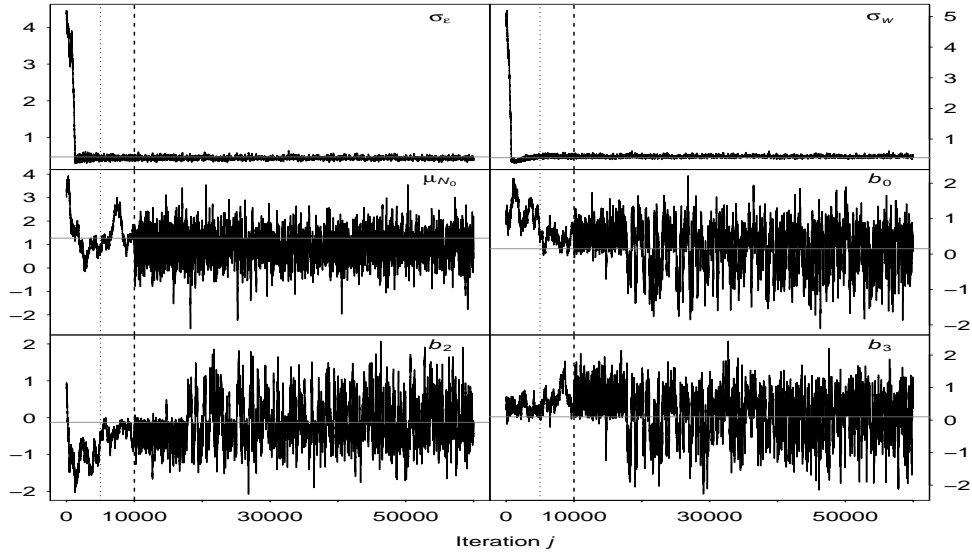}}
\caption{\footnotesize{Trace plots of the sample paths for the marginal Markov chain parameters $\sigma^2_{\epsilon}, \sigma^2_{w}, N_0, b_0, b_2, b_3$ based on $L=5000$ with STAGE 1: the first 5,000 samples from the annealed stage (to the left of the vertical dotted line); STAGE 2: the samples from 5,001 to 10,000 involve the non-adaptive PMCMC stage (between the dotted and dashed lines) and; STAGE 3: the samples from 10,001 to 60,000.}}
\label{fig3:TraceThetaLogistic}
\end{figure}

Figure \ref{fig3:TraceThetaLogistic} shows the trace plots of the Markov chain sample paths for the static parameters $\sigma^2_{\epsilon}, \sigma^2_{w},$ $N_0, b_0, b_2, b_3$. The first 5,000 samples demonstrate that the annealing stage can handle initializations of the Markov chain far from the true parameter values used to generate the data. The second stage, from samples 5,000 to 10,000, demonstrates the slow mixing performance of the untuned non-adaptive chain. Finally the samples from 10,000 to 60,000 clearly show the significant improvement of including an adaption stage in the AdPMCMC algorithm.

\begin{figure}[!ht]
\centerline{\includegraphics[width=0.8\textwidth, height=8cm]{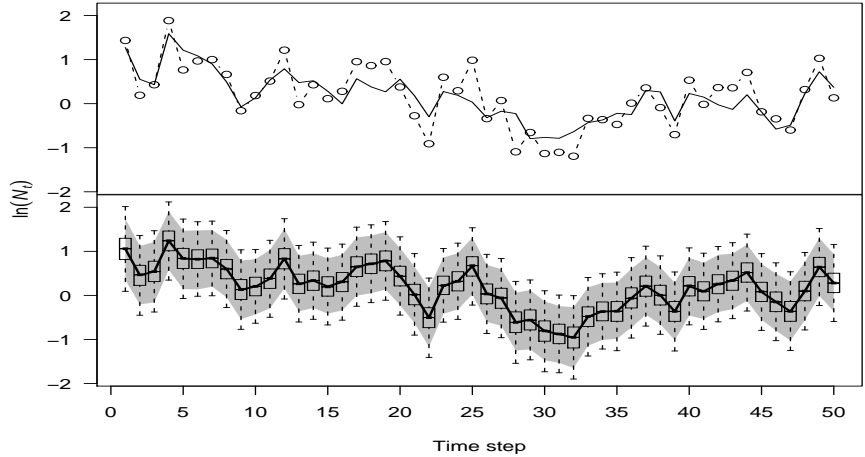}}
\caption{\footnotesize{TOP PANEL: The true latent process (solid line) and the observations (dashed line). BOTTOM PANEL: Box plots for the path space estimates $\{[N_{1:T}](j)\}_{j=1}^J$ obtained from the PMCMC algorithm. In addition, we present the estimated $N^{MMSE}_{1:T}$ as a solid line and the grey colouring presents the $95\%$ posterior predictive intervals for each $N_{t}$.}}
\label{fig4:Data_PathMMSE}
\end{figure}

The top panel of Figure \ref{fig4:Data_PathMMSE} presents the true generated latent process $N_{1:T}^{TRUE}$ and the observations. The bottom panel shows a comparison of the $N_{1:T}^{TRUE}$ versus boxplots of each state element $N_i$ obtained using 50,000 post burn-in samples from the AdPMCMC algorithm. We also show the estimated path space MMSE, $N_{1:T}^{MMSE}$ and the $95\%$ posterior predictive intervals shaded around this MMSE estimate. This plot demonstrates that the MMSE estimate is accurate.

\begin{figure}[!ht]
\centerline{\includegraphics[width=0.7\textwidth, height=8cm]{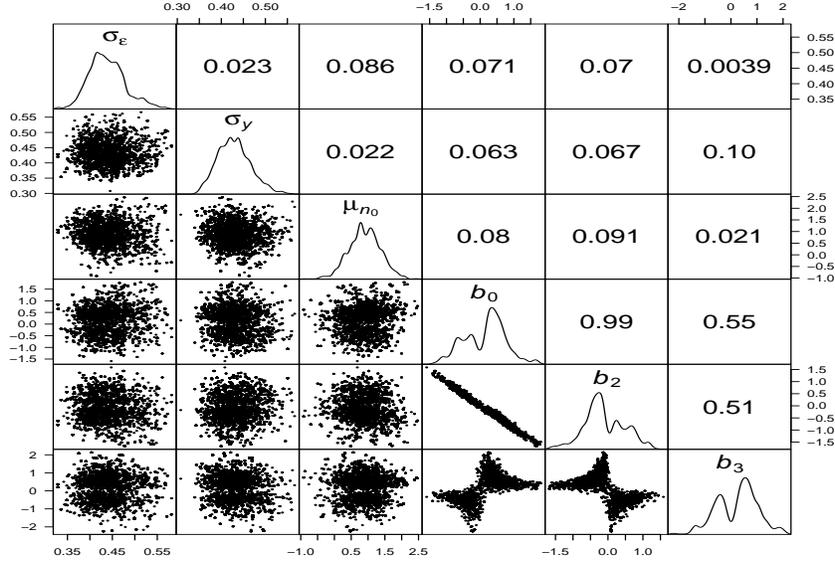}}
\caption{\footnotesize{Scatter plots of the posterior distributions for the static parameters $\bm{\theta}$, this plot also contains the smoothed estimated marginal posterior distributions for each static parameter, followed by the estimated linear posterior correlation coefficient between each static parameter.}}
\label{fig5:boxplots}
\end{figure}

Figure \ref{fig5:boxplots} demonstrates, for $L=5,000$ particles, scatter plots of the pairwise marginal distributions for each static parameter (lower triangular region of the matrix plot), the kernel density estimate of the marginal posterior distribution for each parameter (diagonal of the matrix plot) and the estimated posterior correlation coefficient between the posterior parameters (upper triangular matrix plot).

\begin{figure}[!ht]
\centerline{\includegraphics[width=0.7\textwidth,height=7cm]{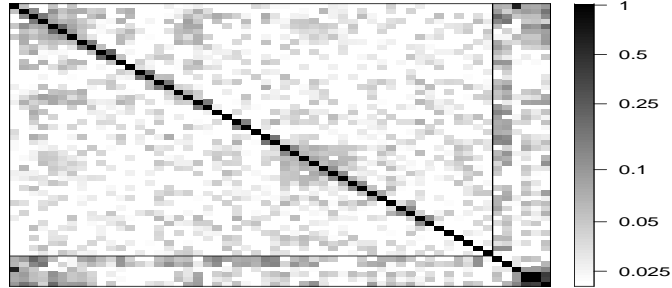}}
\caption{\footnotesize{Heat map in grey scale for the posterior correlation matrix between static parameters and the path space, $N_{1:T}$, given in the first 50 rows and columns, followed by the state parameters in the last six.}}
\label{fig6:ScatterHeat}
\end{figure}

Figure \ref{fig6:ScatterHeat} shows, for $L=5,000$ particles, a heat map for the estimated posterior correlation matrix obtained using samples from the AdPMCMC for the path space and static parameters, $p(\bm{\theta},n_{1:T}|y_{1:T})$.
\vspace{-1cm}
\subsection{Synthetic Data - Mean Square Error (MSE) Analysis}
In this section we study the MSE estimates of the latent process $N_{1:T}$, after integrating out the uncertainty in the static parameters $\bm{\theta}$, for a range of SMC particle counts L. This will illustrate the accuracy of the AdPMCMC estimates of the true underlying process for a given signal to noise ratio. We also derive a recursive expression for the Bayesian Cram\'er-Rao Lower Bound (BCRLB) for each model $M_0$ to $M_4$ as a lower bound comparison. We demonstrate how, for a set of $D$ data sets each of length $T$, the BCRLB can be trivially estimated at no additional computational cost in our model framework, recursively for each time step $t$, via the AdPMCMC algorithm and a modified recursion from \citep{tichavský1998posterior}.

\noindent \textbf{Cram\'er-Rao Lower Bound for the Path Space Proposal in PMCMC} \label{CRLB}\\
To estimate the Cram\'er-Rao Lower Bound in the Bayesian context we do not require that the estimator of interest, in our case $N_{1:T}^{MMSE}$, be unbiased. However we do require that the model is specified such that the following two conditions hold. \\
\noindent \textbf{Condition 1:} for each of the static model parameters $\theta^{(i)} \in [a^{(i)},b^{(i)}]$, the prior model $p(\theta^{(i)})$ satisfies that $\lim_{\theta^{(i)}\rightarrow a^{(i)}}p(\theta^{(i)}) \rightarrow 0$ and $\lim_{\theta^{(i)}\rightarrow b^{(i)}}p(\theta^{(i)}) \rightarrow 0$\\
\noindent \textbf{Condition 2:} The following smoothness properties of the likelihood hold $\theta^{(i)}$,
$$\int \frac{\partial f_{\theta^{(i)}}(y_{1:T})}{\partial \theta^{(i)}} dy_{1:T} = 0$$
Under these conditions we may use the results of \citep{tichavský1998posterior} in which recursive expressions for the BCRLB are derived for general non-linear state space models. We modify these results to integrate out the posterior uncertainty in the joint estimates of the static parameters $\bm{\theta}$. We derive results that perform this marginalization numerically utilizing the existing AdPMCMC framework for each data set.

In particular we estimate for models $M_0$ through to $M_4$ the following mean square error (mse), recursively in time t,
\begin{equation}
\begin{split}
&\int\cdots\int  \left\{\left[N_{1:T}-\widehat{N}_{1:T}\right]\left[N_{1:T}-\widehat{N}_{1:T}\right]^T\right\}
p\left(n_{1:T},y_{1:T}, \bm{\theta}\right) dn_{1:T} dy_{1:T} d \bm{\theta} \\
&=\int\cdots\int \mathbb{E}_{p(n_{1:T},y_{1:T}|\bm{\theta})} \left\{\left[N_{1:T}-\widehat{N}_{1:T}\right]\left[N_{1:T}-\widehat{N}_{1:T}\right]^T\right\}p\left(\bm{\theta}\right) d \bm{\theta},
\end{split}
\label{BCRLBMarg}
\end{equation}
where in this paper we focus on the MMSE estimator $\widehat{N}_{t}=\widehat{N}_{t}^{MMSE}$. The BCRLB provides a lower bound on the MSE used to estimate the path space parameters which correspond in our model to the estimation of the latent process states $N_{1:T}$. We denote the Fisher Information Matrix (FIM), used in the BCRLB, on the path space by $[J_{1:T}\left(N_{1:T}\right)]\left(j\right)$ and marginally by $[J_{t}\left(N_{t}\right)]\left(j\right)$ for time $t$, conditional on the proposed static parameters at iteration $j$ of the PMCMC algorithm. Here we derive an analytic recursive expression for this quantity. For $M_0$ we can get an analytic solution whereas for the other models we will resort to AdPMCMC based online approximations with a novel estimation method based on the particle filter proposal distribution of our AdPMCMC algorithm for each data set.

Conditional on the previous Markov chain state $\left[\bm{\theta}, n_{1:T}\right]\left(j-1\right)$ and the new sampled Markov chain proposal for the static parameters at iteration $j$, $\left[\bm{\theta}\right]\left(j\right) $, we obtain the following modified recursive expression for the FIM based on Eq. (21) in \citep{tichavský1998posterior}:
\begin{equation}
\left[J_t(\widehat{N}_t)\right](j) = \left[D_{t-1}^{22}(\widehat{N}_t)\right](j) -\left[D_{t-1}^{21}(\widehat{N}_t)\right](j) \left(\left[J_{t-1}(\widehat{N}_t)\right](j)+\left[D_{t-1}^{11}(\widehat{N}_t)\right](j)\right)^{-1}\left[D_{t-1}^{12}(\widehat{N}_t)\right](j),
\label{BCRLBRecursion}
\end{equation}
where we obtain the following matrix decompositions of our system model, via Eqs. (34-36) of \\ \citep{tichavský1998posterior} under the model assumptions of additive Gaussian process and observation noise (note derivatives here are taken under the log transformed models for $N_t$ as specified in the additive Gaussian error SSMs in Section \ref{stateEqn}): {\small{
\begin{equation*}
\begin{split}
\left[J_0(\widehat{N}_t)\right](j) &= -\mathbb{E} \left[\bold \nabla_{\log N_{0}}\left\{\bold \nabla_{\log N_{0}} \log p\left(N_0\right)
\right\}^T \right];\\
\left[D_{t-1}^{11}\right](j)  &= -\mathbb{E} \left[\bold \nabla_{\log N_{t-1}}\left\{\bold \nabla_{\log N_{t-1}} \log p\left(N_t|N_{t-1}\right)
\right\}^T \right] = \mathbb{E} \left\{ \left[\nabla_{\log N_{t-1}}f\left(N_{t-1};\bm{\theta}\right)\right]Q_{t-1}^{-1}\left[\nabla_{\log N_{t-1}}f\left(N_{t-1};\bm{\theta}\right)\right]^T\right\};\\
\left[D_{t-1}^{12}\right](j) &= \left[D_{t-1}^{21}\right](j) = -\mathbb{E} \left[\bold \nabla_{ \log N_{t}}\left\{\bold \nabla_{ \log N_{t-1}} \log p\left(N_t|N_{t-1}\right)\right\}^T \right] = -\mathbb{E} \left[\bold \nabla_{ \log N_{t-1}} f\left(N_{t-1};\bm{\theta}\right)\right]Q_{t-1}^{-1};\\
\left[D_{t-1}^{22}\right](j)  &= -\mathbb{E} \left[\bold \nabla_{ \log N_{t}}\left\{\bold \nabla_{ \log N_{t}} \log p\left(N_t|N_{t-1}\right)
\right\}^T \right] + -\mathbb{E} \left[\bold \nabla_{ \log N_{t}}\left\{\bold \nabla_{ \log N_{t}} \log p\left(y_t|N_{t}\right)
\right\}^T \right]\\
&= Q_{t-1}^{-1} + \mathbb{E}\left\{\left[\mathbf \nabla_{ \log N_{t}} h\left(N_{t};\bm{\theta}\right) \right]R_{t}^{-1}\left[\bold \nabla_{ \log N_{t}} h\left(N_{t};\bm{\theta}\right) \right]^T\right\}
\end{split}
\end{equation*} }}
where $f\left(N_{t-1};\bm{\theta}\right)$ is the state model with process noise covariance $Q_t$ and $h\left(N_{t};\bm{\theta}\right)$ is the observation model with observation noise covariance $R_t$. Next we derive these quantities for each model, summarised in Table \ref{tab:CRLBDerivation} where the expectations terms $\left[J_0(\widehat{N}_0)\right](j) = \left[\frac{1}{\sigma_{\epsilon}^2}\right](j)$ and $\left[D_{t-1}^{22}\right](j) = \left[\frac{1}{\sigma_{\epsilon}^2} + \frac{1}{\sigma_w^2}\right](j)$ are common to all models and evaluate as shown for iteration $j$ of the AdPMCMC algorithm.

\begin{table}
\caption{\label{tab:CRLBDerivation} Derivation of the BCRLB for each model.}
\centering
\fbox{
\begin{tabular}{*{3}{c}} 
\textbf{Model} & $\left[D_{t-1}^{11}\right](j)$ & $\left[D_{t-1}^{12}\right](j)$ \\ \hline
$M_0$ & $\frac{1}{\sigma_{\epsilon}^2}$ & $-\frac{1}{\sigma_{\epsilon}^2}$ \\
$M_1$ & $\mathbb{E}\left[\left(1 + \exp(N_{t-1})\right)^2\frac{1}{\sigma_{\epsilon}^2}\right]$ & $-\mathbb{E}\left[\left(1 + \exp(N_{t-1})\right)\frac{1}{\sigma_{\epsilon}^2}\right]$\\
$M_2$ & $\mathbb{E}\left[\left(1 + b_2 b_3 \exp\left(b_3 N_{t-1} \right)\right)^2\frac{1}{\sigma_{\epsilon}^2}\right]$ & $-\mathbb{E}\left[\left(1 + b_2 b_3 \exp\left(b_3 N_{t-1} \right)\right)\frac{1}{\sigma_{\epsilon}^2}\right]$ \\
$M_3$ & $\mathbb{E}\left[\left(2 - \frac{\exp\left(N_{t-1}\right)}{\left(b_4 +  \exp\left(N_{t-1}\right) \right)} + b_1 \exp\left(N_{t-1}\right)\right)^2 \frac{1}{\sigma_{\epsilon}^2}\right]$ & $-\mathbb{E}\left[\left(2 - \frac{\exp\left(N_{t-1}\right)}{\left(b_4 + \exp\left(N_{t-1}\right) \right)} + b_1 \exp\left(N_{t-1}\right)\right)\frac{1}{\sigma_{\epsilon}^2}\right]$ \\
$M_4$ & $\mathbb{E}\left[\left(1 + b_6\exp\left(N_{t-1}\right) + 2b_7\exp\left(2N_{t-1}\right)\right)^2\frac{1}{\sigma_{\epsilon}^2}\right]$ & $-\mathbb{E}\left[\left(1 + b_6\exp\left(N_{t-1}\right) + 2b_7\exp\left(2N_{t-1}\right) \right)\frac{1}{\sigma_{\epsilon}^2}\right]$ \\ \hline
\end{tabular}}
\end{table}

\textbf{Remark 1:} \textit{The key point about utilising this recursive evaluation for the FIM matrix is that in the majority of cases one can not evaluate the required expectations in Table \ref{tab:CRLBDerivation} analytically. However, since we are constructing a particle filter proposal distribution for the AdPMCMC algorithm to target the filtering distribution $p\left(n_t|y_{1:T},[\bm{\theta}](j)\right)$ we can use this particle estimate to evaluate the expectations at each iteration $t$. It is important to note that this recursion avoids calculating the expectations using the entire empirical estimate of the path space, and only requires the marginal filter density estimates for each data set, which will not suffer from degeneracy as a path space emprical estimate would. Hence, for example we approximate the expectation at each time recursion t with $\mathbb{E}\left[\left(1 + \exp(N_{t-1})\right)^2\frac{1}{\sigma_{\epsilon}^2}\right]\approx \sum_{i=1}^L \left[W_t \left(1 + \exp(N_{t-1})\right)^2\frac{1}{\sigma_{\epsilon}^2}\right](j,i)$ using the current Markov chain realisation of the parameters $[\bm{\theta}](j)$ and the particle estimate $[N_{t-1}](j,i)$ for the $i$-th particle at iteration $j$ of the AdPMCMC algorithm, for each data set.}

\textbf{Remark 2:} \textit{We can estimate accurately the BCRLB at each stage of the filter whilst simultaneously integrating out the static parameters $\bm{\theta}$. In addition we note that for the Exponential model ($M_0$), the BCRLB is analytic and optimal in this recursion and is given by the classic information filter, see \citep{harvey1991forecasting}. That is, in the AdPMCMC algorithm for $M_0$, the particle filtering proposal can be replaced via Rao-Blackwellization with the Kalman filter to obtain the marginal likelihood evaluations \citep{carter1994gibbs} for this special case of the linear Gaussian system.  
}

Table \ref{tab:MSE_results} presents simulation results for models $M_0$ through to $M_4$. In particular using the estimates obtained by the AdPMCMC algorithm for the Bayesian MMSE estimates in Equation (\ref{BayesianEstimators}) for the latent process $N_{1:T}$ we estimate the following mean square error (mse) quantity for each model
\begin{equation}
\begin{split}
\frac{1}{T} \int& \sum_{t=1}^{T}\mathbb{E}_{p(n_{1:T},y_{1:T}|\bm{\theta})} \left\{\left[\widehat{N}_{t}^{MMSE} - n_{t}^{TRUE}\right]\left[\widehat{N}_{t}^{MMSE} - n_{t}^{TRUE}\right]^T|\bm{\theta}\right\} p\left(\bm{\theta}\right) d\bm{\theta} \\
& \approx \frac{1}{JT}\sum_{j=1}^{J}\sum_{t=1}^{T} \mathbb{E}\left\{\left[ \left[\widehat{N}_{t}^{MMSE}\right](j) - n_{t}^{TRUE}\right]\left[\left[\widehat{N}_{t}^{MMSE}\right](j) - n_{t}^{TRUE}\right]^T|\left[\bm{\theta}\right](j)\right\}\\
& = \frac{1}{JT}\sum_{j=1}^{J}\sum_{t=1}^{T} \mathbb{E}\left\{\left[\frac{1}{L}\sum_{i=1}^{L}[W_t N_t](j,i) - n_{t}^{TRUE}\right]\left[\frac{1}{L}\sum_{i=1}^{L}[W_t N_t](j,i) - n_{t}^{TRUE}\right]^T|\left[\bm{\theta}\right](j)\right\}
\end{split}
\label{EstimatedMarginalMSE}
\end{equation}
where $\bm{\theta}$ denotes the static parameters. We then obtain an average of this estimate over $D=20$ independent data realizations and report this as the average MSE for the latent state process estimate for each model and in (brackets) the standard deviation of the MSE estimate is reported. We then compare these average MSE estimates to the BCRLB estimates derived above to provide a performance comparison of our methodology as a function of the number of particles, $L$. 

The results are presented here for each model with $L\in\left\{20,100,500\right\}$ particles, $T=50$ and $50,000$ iterations post burn-in ($J=50,000$). We report the results for 20 blocks per AdPMCMC chain in calculation of the estimated Root Mean Square Error for each chain. However, we also explored varying the number of blocks between (10, 20, 40) in the RMSE calculations which we found to have only a minor effect on the average MSE estimate. 

We present two tables of results, the first in Table \ref{tab:MSE_results} utilises data generated from model settings which reflect typical observation and process noise variance levels and the second in Table \ref{tab:MSE_noise} presents results in which the noise variance is increased by a factor of two and four. Each data set is generated from the same model parameters in each of the models $M_0$ to $M_4$ for comparison purposes. 

\begin{table}
\caption{\label{tab:MSE_results}Average RMSE for $N_{1:T}^{MMSE}$ for 20 independent data realizations (T = 50).}
\centering
\fbox{
\begin{tabular}{*{6}{c}}
\multicolumn{6}{c}{Average Estimated Root Mean Square Error (20 blocks, J=50,000)} \\ \hline
L - number of particles & $M_0$ & $M_1$ & $M_2$ & $M_3$ & $M_4$ \\ \hline
20 & 0.31 (0.03) & 0.32 (0.04) & 0.31 (0.03) & 0.33 (0.03) & 0.32 (0.03)\\
100& 0.31 (0.04) & 0.31 (0.03) & 0.30 (0.03) & 0.31 (0.04) & 0.32 (0.04)\\
500& 0.31 (0.03) & 0.32 (0.04) & 0.31 (0.04) & 0.31 (0.04) & 0.30 (0.04)\\ 
\multicolumn{6}{c}{Average Estimated BCRLB} \\ \hline
500 & 0.34 (0.01) & 0.38 (0.01) & 0.34 (0.01) & 0.33 (0.01) & 0.35(0.02)\\  
\end{tabular}}
\end{table}

The results demonstrate the accuracy of our methodology since the MMSE estimates obtained from the simulations are not (statistically) significantly different from the estimated average BCRLB. This provides confidence in the AdPMCMC methodology. In addition we see that increasing the noise variance of the observations, results in a relatively larger increase in the estimated RMSE compared to the estimated BCRLB. In other words as the signal-to-noise ratio decreases, the estimator accuracy also decreases, therefore producing a larger difference between the average BCRLB and the estimated average RMSE. We see this effect in Table \ref{tab:MSE_noise}, though it is not overly pronounced. The results in most models at high noise settings no longer achieve the BCRLB on average, though they are still reasonably close. 

\begin{table}
\caption{\label{tab:MSE_noise}Average RMSE for $N_{1:T}^{MMSE}$ for 20 independent data realizations (T = 50).}
\centering
\fbox{
\begin{tabular}{*{6}{c}}
\multicolumn{6}{c}{Average Estimated Root Mean Square Error (20 blocks, J=50,000, L = 500)} \\ \hline
$\sigma_{w}^2$ & $M_0$ & $M_1$ & $M_2$ & $M_3$ & $M_4$ \\ \hline
$2 \times$ & 0.40 (0.05) & 0.39 (0.05) & 0.39 (0.03) & 0.39 (0.05) & 0.39 (0.06) \\
$4 \times$ & 0.50 (0.07) & 0.50 (0.05) & 0.51 (0.05) & 0.49 (0.07) & 0.51 (0.07) \\
\multicolumn{6}{c}{Average Estimated BCRLB} \\ \hline
$2 \times$ & 0.38 (0.02) & 0.42 (0.02) & 0.36 (0.02) & 0.35 (0.02) & 0.38 (0.02) \\ 
$4 \times$ & 0.43 (0.03) & 0.47 (0.03) & 0.41 (0.03) & 0.38 (0.02) & 0.43 (0.03) \\ 
\end{tabular}}
\end{table}

\subsection{Bayes Factors Model Selection Assessment.}
The ideal case for choosing between posterior models occurs when integration of the posterior (with respect to the parameters) is possible. For two models M$_{i}$ and M$_{j}$, parameterized by $\theta _{1:k} $ and $\alpha _{1:j} $ respectively, the posterior odds ratio of M$_{i}$ over M$_{j}$ is given by
\begin{equation*}
\frac{p\left( {y_{1:T} \vert M_i } \right)p(M_i )}{p\left(
{y_{1:T} \vert M_j } \right)p(M_j )} = \frac{p(M_i )\int {p\left(
{y_{1:T} \vert \theta _{1:k} ,M_i } \right)} p\left( {\theta
_{1:k} \vert M_i } \right)d\theta _{1:k} }{p(M_j )\int {p\left(
{y_{1:T} \vert \alpha _{1:j} ,M_j } \right)} p\left( {\alpha
_{1:j} \vert M_j } \right)d\alpha _{1:j} } = \frac{p(M_i )}{p(M_j
)}BF_{ij}
\end{equation*}
\noindent which is the ratio of posterior model probabilities, having observed the data, where $p(M_{i})$ is the prior probability of model $M_{i}$ and $BF_{ij}$ is the Bayes factor. This quantity is the Bayesian version of a likelihood ratio test, and a value greater than one indicates that model $M_{i}$ is more likely than model $M_{j}$ (given the observed data, prior beliefs and choice of the two models).

\begin{figure}[!ht]
\centerline{\includegraphics[width=0.8\textwidth, height=6cm]{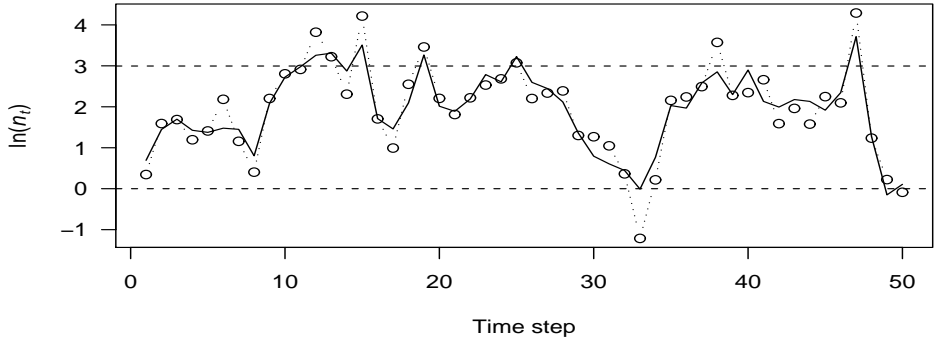}}
\caption{\footnotesize{A representative data realization for the BF analysis generated from model $M_4$ the Flex-Ricker model. The upper dashed line is the carrying capacity (K) and the lower dashed line is the Allee threshold (C). }}
\label{fig8:BFSynth}
\end{figure}

To demonstrate the performance of the AdPMCMC in this context we generated 10 data sets using the flexible-Allee model, $M_4$ each with the same model parameters. Then we estimated each of the models $M_0$ through to $M_4$ using this data, and calculated estimates of the Bayes factors. A representative data realization for the parameters of $M_4$ chosen in this simulation is presented in Figure \ref{fig8:BFSynth}. 
In this example we considered $ N_0 = \ln(2), b_5 = -0.05, b_6 = 0.0525, b_7 = -0.0025$ with process and observation noise variances drawn randomly from the priors as detailed in Section 3.1 . These model parameters were selected to generate population trajectories that fluctuate between the population carrying capacity $K$ and the Allee threshold $C$. This makes for a challenging model selection scenario and should result in ambiguity over the actual true model, as all models we consider can potentially capture this form of population growth behaviour in the presence of the high process and observation noise used here. The results in Table \ref{tab:AveLNBF} confirm that for any given realization of the data from M4 with these model parameters, we see switching between the most plausible model to explain the particular realization. We see from the Bayes Factors that there is a strong ambiguity between the true model, the flexible Allee $M_4$ used to generate the data and the Ricker model $M_1$. The reason for this is that the process and observation noise obscures the Allee effect but the signal of negative density dependence remains, and this is most parsimoniously represented by the Ricker model $M_1$.

This synthetic example highlights the challenges faced in choosing between models, when realistic levels of observation and latent process noise are present in the population counts data. It also emphasises the importance of efficient posterior MCMC samplers for each model to ensure that the Bayes Factors are accurate. These results also highlight that observation noise and process noise modelling can be critical when determining the presence of density dependent mortality or Allee effects in real data sets.
\begin{table}
\caption{\label{tab:AveLNBF}Bayes Factors (rounded to integers) with true model as $M_4$ and (T = 50, L=500, J=50k). }
\centering
\fbox{
\begin{tabular}{*{11}{c}}
Data Set & $BF_{01}$ & $BF_{02}$ & $BF_{03}$& $BF_{04}$ & $BF_{12}$ & $BF_{13}$ & $BF_{14}$ & $BF_{23}$ & $BF_{24}$ & $BF_{34}$	\\ \hline
1 & 0 & 1 & 3 & 0 & 32 & 185 & 6 & 6 & 0 & 0 \\
2 & 0 & 0 & 0 & 0 & 411 & 24 & 3 & 0 & 0 & 0 \\
3 & 40 & 31 & 11873 & 0 & 1 & 296 & 0 & 379 & 0 & 0 \\
4 & 0 & 0 & 0 & 0 & 18986 & 62 & 5 & 0 & 0 & 0 \\
5 & 0 & 0 & 0 & 0 & 2 & 66 & 3 & 27 & 1 & 0 \\
6 & 0 & 0 & 0 & 0 & 8 & 15 & 5 & 2 & 1 & 0 \\
7 & 0 & 0 & 0 & 0 & 65903 & 44 & 5 & 0 & 0 & 0 \\
8 & 0 & 0 & 0 & 0 & 27 & 97 & 137 & 4 & 5 & 1 \\
9 & 0 & 0 & 0 & 0 & 697 & 98 & 4 & 0 & 0 & 0 \\
10 & 0 & 0 & 0 & 0 & 75 & 5229 & 3 & 70 & 0 & 0 \\
\end{tabular}}
\end{table}

When the standard deviations of both the process and observation noise were decreased by an order of magnitude, then the model selection exercise produces Bayes Factors that identify the correct model, (Table \ref{tab:AveLNBF_obs}). In this reduced noise case, the Bayes Factors for all models (other than $M_4$) versus model $M_4$ were less than one. Table \ref{tab:AveLNBF_Noise} explores the effect of observation and process noise on the model selection under different signal to noise ratios. We focus on the Bayes factors for model $M_1$ versus $M_4$, as these two models are most likely to be ambiguous in this case as they are each capable of capturing an Allee effect to varying degrees.


\begin{table}
\caption{\label{tab:AveLNBF_obs}Bayes Factors (rounded to integers) with true model as $M_4$ and (T = 50, L = 500, J = 50k). Observation noise variance and process noise variance both decreased by two orders of magnitude.}
\centering
\fbox{
\begin{tabular}{*{11}{c}}
Data Set & $BF_{01}$ & $BF_{02}$ & $BF_{03}$& $BF_{04}$ & $BF_{12}$ & $BF_{13}$ & $BF_{14}$ & $BF_{23}$ & $BF_{24}$ & $BF_{34}$	\\ \hline
1 & 0 & 2 & 0 & 0 & 68 & 13 & 0 & 0 & 0 & 0 \\
2 & 0 & 0 & 0 & 0 & 41 & 1 & 0 & 0 & 0 & 0 \\
3 & 0 & 3 & 0 & 0 & 2207 & 2 & 1 & 0 & 0 & 0 \\
4 & 0 & 3 & 0 & 0 & 275 & 4 & 0 & 0 & 0 & 0 \\
5 & 0 & 0 & 0 & 0 & 83 & 2 & 0 & 0 & 0 & 0 \\
6 & 0 & 2 & 0 & 0 & 65 & 0 & 0 & 0 & 0 & 0 \\
7 & 0 & 2 & 0 & 0 & 33 & 1 & 0 & 0 & 0 & 0 \\
8 & 0 & 2 & 0 & 0 & 144 & 1 & 0 & 0 & 0 & 0 \\
9 & 0 & 7 & 0 & 0 & 39 & 1 & 1 & 0 & 0 & 1 \\
10 & 0 & 1 & 0 & 0 & 116 & 2 & 0 & 0 & 0 & 0 
\end{tabular}}
\end{table}


\begin{table}
\caption{\label{tab:AveLNBF_Noise}Bayes Factor $BF_{14}$ (rounded to integers) with true model as $M_4$ and (T = 50, L=500, J=50k). }
\centering
\fbox{
\begin{tabular}{*{11}{c}}
\multicolumn{11}{c}{Data Set} \\ 
Noise Levels & 1 & 2 & 3 & 4 & 5 & 6 & 7 & 8 & 9 & 10 \\ \hline
$\sigma^2_{\epsilon,w} \sim IG(T/2, (T - 2)/10)$  &6 &3 &0 &5 &3 &5 &5 &137 &4 &3\\
$\sigma_{\epsilon}^2/10$  &0 &7 &0 &0 &10 &10 &0 &1 &3 &0\\
$\sigma_w^2/10$ 	&6 &2 &0 &1 &121 &0 &2 &7 &1 &0\\
$\sigma_{\epsilon, w}^2/10$ 	&0 &2 &0 &0 &0 &0 &0 &1 &0 &0\\
$\sigma_{epsilon}^2/100$  &0 &1 &0 &1 &4 &0 &0 &0 &1 &0\\
$\sigma_w^2/100$  &10 &1 &0 &39 &1 &0 &7 &2 &0 &2\\
$\sigma_{\epsilon, w}^2/100$  &0 &0 &1 &0 &0 &0 &0 &0 &1 &0
\end{tabular}}
\end{table}

The results in Table \ref{tab:AveLNBF_Noise} demonstrate several important points relating to model selection in the presence of differing severities of process and observation noise. In low process error settings we see that in the majority of cases, the correct model has strong evidence for its selection. Secondly, in the case in which the observation noise is significantly reduced, we see a reduction in the evidence for the incorrect model $M_1$. However, this is not as marked as when the process error is decreased, indicating that perhaps the process error may have a stronger influence on the ability to distinguish the two closely related models. Finally, we confirm that for large reductions in both process and observation noise, there is clear selection of the appropriate model $M_4$.

\subsection{Real Data Analysis}
In the following section we present analysis of two real data sets. For each data set, we extended the length of the AdPMCMC chains so that the mixing results in a Geweke times series diagnostic approximate Z-score in the interval [-2,2] for all parameters of all models. This ensures accurate estimation of the Bayes factors.

We study the nutria data from \citep{gosling1981dynamics}. Nutria are a widespread invasive species with populations now established around the world \citep{Woods1992}. This data is available as data set 9833 in the Global Population Database \citep{NERC1999}, and presents a time series of female nutria abundance in East Anglia at monthly intervals, obtained by retrospective census for a feral population.

This data set is of interest because \citep{Drake2005} fit models $M_0 \ldots M_3$ without observation error, and found that the AIC selected the strong Allee model $M_3$ as the best overall model whereas BIC selected the Ricker model $M_1$ as the best overall model. Here we revisit this data set, include observation error, add an additional model ($M_4$) that allows for both strong and weak Allee effects, and use Bayes factors for model selection.

\begin{figure}[ht!]
\centerline{\includegraphics[width=0.8\textwidth,height=8cm]{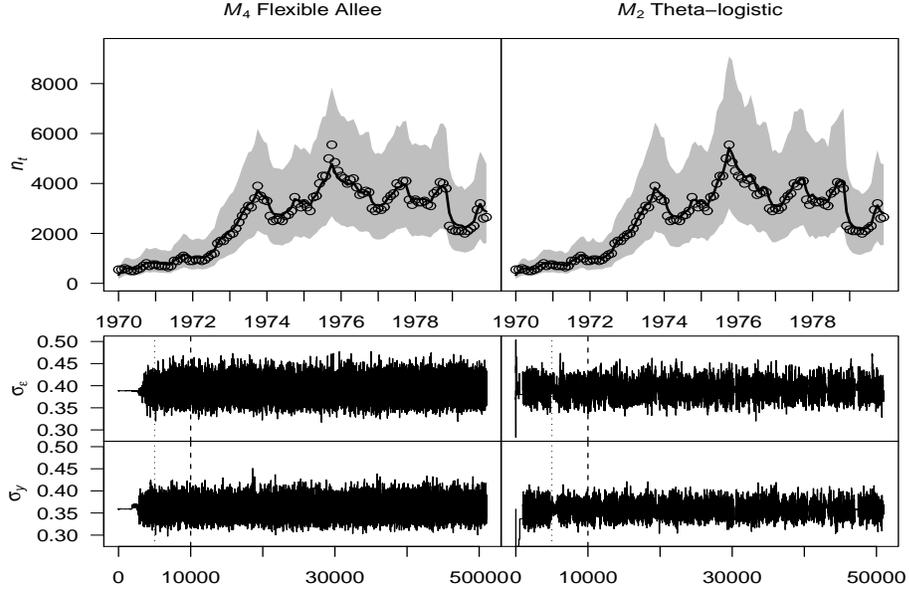}}
\caption{\footnotesize{Nutria data analysis. Left plots - model $M_4$ and Right plots - model $M_2$. Top panel - estimates of the MMSE for the path space (solid lines), observations from data (points) and posterior 95\% predicitive interval (shaded). Lower panels - trace plots of noise standard deviations for observation and state process.}}
\label{fig8a:NutriaResult}
\end{figure}

In Figure \ref{fig8a:NutriaResult} we present the two most plausible models for the nutria data, according to the Bayes factor analysis in Table \ref{tab:NutriaLNBF} with the following settings: T = 120, L=500, J = 350,000 post burn-in, (burn-in = 155,000 and annealing = 5,000). In this analysis, the default scalings proposed by \citep{roberts2009examples} for the non-adaptive component in Algorithm 2 proved suboptimal for models $M_{1, 3, 4}$. We subsequently scaled down the non-adaptive components in these models by factors of $10^{-3}$ to $10^{-7}$ and made a similar adjustment in the annealing phase. The MMSE estimates for the path space $N_{1:T}^{MMSE}$, thinned by $10\%$, are shown for models $M_4$ and $M_2$, together with the unthinned Markov chain trace plots of the noise and process model variances. The solid lines are the MMSE of the path space and the points are observations with the gray shading corresponding to the posterior 95\% credibility intervals.

Our findings demonstrate posterior evidence for the selection of model $M_4$, over all other models, in the presence of jointly estimated observation and state noise. With the selection of model $M_4$, it is interesting to comment on the MMSE estimate for the model parameters from this fit. There is a small probability ($<0.01$) that the only stable equilibrium is zero; that is, there is no stable positive equilibrium $K$ and the population will go extinct. Conditional on the existence of a positive per capita population growth rate, that is positive for some population density, the estimate for the Allee threshold $C$ in model $M_4$ is -2484 (mean) $\left[-38814, 796\right]$ (0.95 CI) relative to a carrying capacity $K = 3403 \left[2369, 4437\right]$. This suggests that a weak Allee effect is present in the nutria population.

If we ignore the new model $M_4$ , the Bayes factors (Table \ref{tab:NutriaLNBF}) identify model $M_1$, which has no Allee effect, as the best fitting model, over model $M_3$, which only has a strong Allee effect. In contrast, the results of \citet{Drake2005} are ambiguous, where AIC results suggest that $M_3$ is the best model but BIC suggests $M_1$ is the best. \citet{Drake2005} notes that the differences in both AIC and BIC results are minor and do not provide compelling support for one model over the other. These results show the utility of considering models (such as $M_4$) that allow for a weak Allee effect

The second real data set we analyse is a time series of a population of sparrowhawks (\emph{Accipiter nisus}) in south Scotland. This is a well studied population \citep{Newton1997} that shows evidence of density dependence \citep{Newton1986}. This data set (number 6575 in the Global Population Database \citep{NERC1999}) is of interest because \citep{Polansky2009} showed that the likelihood surface of the theta-logistic model applied to this data, with both process and observation noise, has strong ridges and multiple modes. \citep{Polansky2009} calculate the maximum likelihood estimates of the theta-logistic's parameters and warn that its complex likelihood surface may derail methods of model comparison. They calculated that the parameter controlling the form of density dependence (equivalent to parameter $b_3$ in $M_2$) equals -4.83 at the global maximum of the likelihood surface. A second local maximum was found when $b_3 = 0.04$.

\begin{figure}[ht!]
\centerline{\includegraphics[width=1\textwidth,height=10cm]{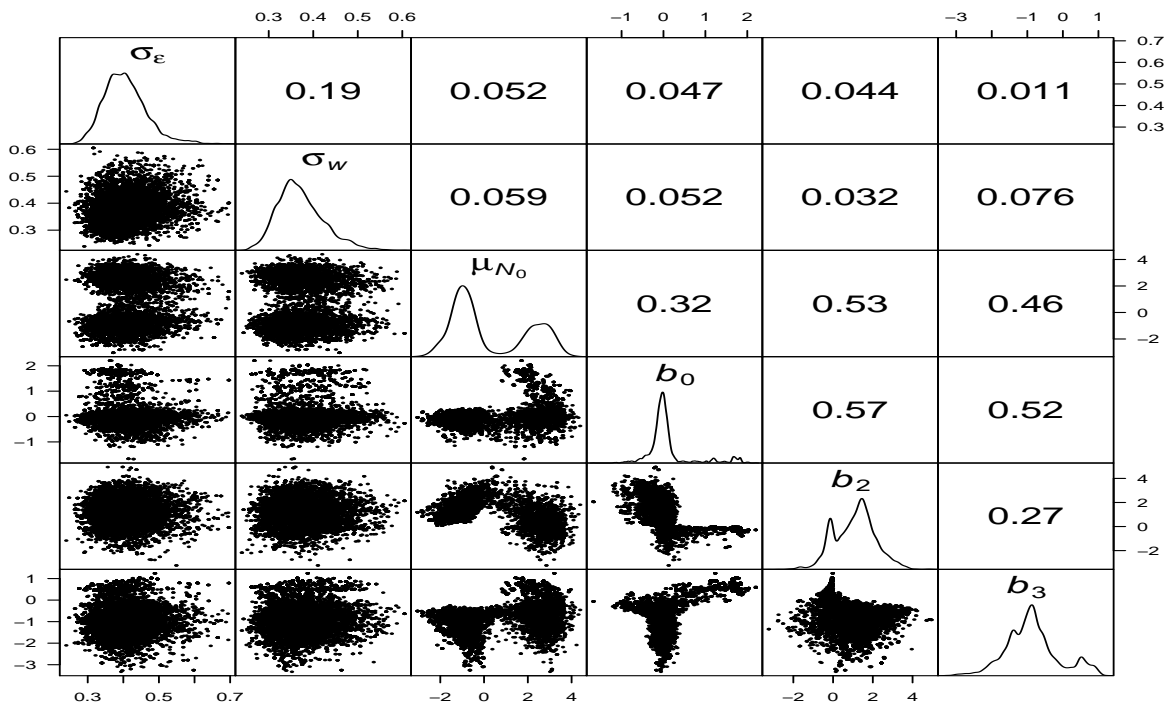}}
\caption{\footnotesize{Scatter plots of the posterior distributions for the static parameters $\bm{\theta}$ for A. nisus data set under model $M_2$. This plot also contains the smoothed estimated marginal posterior distributions for each static parameter, followed by the estimated linear posterior correlation between each static parameter.}}
\label{fig9:AnisusResults}
\end{figure}

The Bayes factors (Table \ref{tab:AnisusLNBF}) with algorithm settings (T = 18, L = 500, J = 150,000 post burn-in, burn-in = 50,000 and annealing = 5,000), however, suggest that the Ricker model $M_1$ is the best fitting model., This is equivalent to the theta-logistic $M_2$ with $b_3 = 1$. In our analysis the theta-logistic model is the worst of the five models. Note also that the AdPMCMC sampler mixes suitably well over the complicated multimodal posterior support (Figure \ref{fig9:AnisusResults}). Here we have again used the Geweke diagnostic to determine the number of post burn-in samples, but we recognise that we may require a larger number of samples in order to accurately estimate the weighting or proportion of posterior mass in each of the two modes. The fact remains that the sampler is able to mix over the two modes, enabling model selection via Bayes factors even in the presence of ridges and multiple modes in the likelihood surface, assuming that the sampler mixes sufficiently well over the posterior support. 

\begin{table}
\caption{\label{tab:NutriaLNBF} Bayes Factors (rounded) for nutria.}
\centering
\fbox{
\begin{tabular}{*{6}{c}}
Model 			& $M_0$ 	& $M_1$ & $M_2$ & $M_3$ & $M_4$ \\ \hline
$BF_{0.}$ 	& 1 			& 0			& 7			& 0 		& 0 \\
$BF_{1.}$ 	& 6432		& 1 		& 44366	& 14 		& 0 \\
$BF_{2.}$ 	& 0				& 0			& 1 		& 0 		& 0 \\
$BF_{3.}$ 	& 450			& 0		  & 3104 	& 1 		& 0 \\ 
$BF_{4.}$ 	& 32229 	& 5 		& 222302& 72		& 1 
\end{tabular}}
\end{table}

\begin{table}
\caption{\label{tab:AnisusLNBF} Bayes Factors (rounded) for \textit{A. nisus}. }
\centering
\fbox{
\begin{tabular}{*{6}{c}}
Model 			& $M_0$ 	& $M_1$ & $M_2$ & $M_3$ & $M_4$ \\ \hline
$BF_{0.}$ 	& 1 			& 0			& 5			& 0 		& 0 \\
$BF_{1.}$ 	& 22			& 1 		& 118		& 10 		& 6 \\
$BF_{2.}$ 	& 0				& 0			& 1 		& 0 		& 0 \\
$BF_{3.}$ 	& 2				& 0		  & 12 		& 1 		& 1 \\ 
$BF_{4.}$ 	& 4 			& 0 		& 21 		& 2			& 1 \\ 
\end{tabular}}
\end{table}

\section{Discussion}

This paper presents a novel technical development: a sophisticated sampling methodology based on Adaptive MCMC \citep{roberts2009examples} and Particle MCMC \citep{AndrieuDoucetHolenstein2010}; that enables a novel application: an efficient and robust approach to jointly estimate observation and process noise with the latent states and static parameters in Bayesian nonlinear state space models of population dynamics. We developed a novel adaptive version of the PMCMC algorithm and demonstrated its performance on several challenging high dimensional posterior models. These models were based on population dynamic models that are cited widely in the literature. We have also placed the problem of statistical inference and model selection in a more realistic setting by accounting for, and estimating, process error and observation noise jointly with the model parameters and latent population process,  without the need for time-consuming tuning and algorithm design. We believe that the novel sampling framework presented here can be easily generalised to different ecological state space models.

The AdPMCMC sampling methodology we developed is more sophisticated than the standard MCMC algorithms currently employed in ecological settings \citep{Millar2000, Clark2004, Ward2006} in which the equations describing the latent population dynamics are often highly non-linear. In addition, there is typically a strong correlation between the model parameters and the latent process. These two factors will make the mixing properties of basic MCMC or block-Gibbs sampling algorithms, highly inefficient in a time series setting, see \citep{AndrieuDoucetHolenstein2010} and \citep{NevatPetersDoucet2009} for discussion.

To understand this, consider the simple sampling framework in which the posterior distribution for model $M_i$, denoted $p(\bm{\theta},n_{1:T}|y_{1:T},M_i)$ is sampled in the following block Metropolis-Hastings within Gibbs framework, where the vector of latent states is split into $k$ block of length $k \tau= T$,
\begin{equation*}
\begin{split}
\text{iteration j: } &\bm{\Theta} \sim p(\bm{\theta}|n_{1:T},y_{1:T},M_i)\\
\text{iteration j: } & N_{1:\tau} \sim p(n_{1:\tau}|\bm{\theta},n_{\tau+1:T},y_{1:T},M_i)\\
\vdots\\
\text{iteration j: } & N_{(k-1)\tau+1:T} \sim p(n_{(k-1)\tau+1:T}|\bm{\theta},n_{1:(k-1)\tau},y_{1:T},M_i).\\
\end{split}
\end{equation*}

Such a block design is highly inefficient if the parameters of the posterior distribution are correlated, as in the case of the \emph{A.~ nisus} data set (Figure \ref{fig9:AnisusResults}). For moderate sized values of $\tau$ this sampling framework will mix poorly because the Metropolis-Hastings acceptance probabilities will be low irrespective of the proposals that are used. The simplest solution to this problem would be to sample directly from the full conditional distributions in a univariate Gibbs framework. However, even in the case where a single component Gibbs sampler is possible via inversion sampling from the full conditional cdf's for all parameters, the Markov chain will mix very slowly around the support of the posterior. This is especially problematic in high dimensional target posterior distributions, since it requires excessively long Markov chains to achieve samples from the stationary regime. It can also lead to very high autocorrelations in the Markov chain states, with a concomitant impact on the variance of the estimators in Equation \ref{BayesianEstimators}. To avoid slow mixing Markov chains, one must sample from larger blocks of parameters, - \hbox{i.e.} larger $\tau$. However, the design of an optimal proposal distribution for large blocks of parameters is very complicated. The Particle MCMC methodology steps around this problem by approximating the optimal proposal distribution for a large number of parameters via a Sequential Monte Carlo proposal distribution.

Another positive property of the AdPMCMC methodology is evident in the Markov chain sample paths, shown in Figure \ref{fig3:TraceThetaLogistic}. As the AdPMCMC sampler mixes over different modes , the estimated adapted covariance in the static parameter proposal ``adapts'' or ``learns'' on-line to account for the additional mode and allows for more efficient mixing between the two modes of the posterior. This demonstrates the power of the adaptive MCMC methodology when combined within the PMCMC sampler.

Previous meta-analyses of ecological time-series have used model selection techniques to conclude that Allee effects are rare without explicitly accounting for the confounding effects of observation error \citep{Saether1996, Sibly2005, Gregory2010}. Here we used a synthetic dataset to show that moderate process and observation error will lead to ambiguous model selection results in ecological time series that include strong Allee effects.This occurs because the noise masks the full signal of the determinstic density-dependent processes. Our results suggest, however, that negative density dependence is much easier to detect than the Allee effect. Indeed, a parsimonious model that includes negative dependence but omits the Allee effect will often be selected in the presence of observation error. The chance of correctly choosing a model with an Allee effect can be increased by decreasing the observation error. 

Nevertheless, we did find evidence for a weak Allee effect in the nutria data set studied by \citep{Drake2005}, who reported the AIC and BIC model selection in the absence of observation noise. We included observation noise and extended the class of models to include $M_4$, which allows for both strong and weak Allee effects, and we found posterior evidence for model $M_4$ over all other models. Furthermore, we found that the second most likely model to explain the data was $M_1$, which lacks an Allee effect, over model $M_3$, which only has a strong Allee effect. This agrees with the BIC results in the analysis without observation noise of \citep{Drake2005}.

In the A. nisus data set, \citep{Polansky2009} use a numerical approach inspired by \citep{Kitigawa1987} to find multiple modes with local optima that suggests a nonlinear relationship between population density and per capita growth rate. The AdPMCMC sampling methodology demonstrates efficient mixing between these multiple modes. \citep{Kitigawa1987} acknowledges that a grid-based approach will not scale efficiently with increasing dimensions in the latent path space or the static parameters; for instance, multispecies analyses will be precluded. On the other hand, our AdPMCMC approach will scale up to much larger dimensions. If an optimization algorithm is used instead, as in \citep{Polansky2009}, then it becomes difficult to calculate the uncertainty associated with the maximum likelihood estimates. In particular, it is difficult to determine the joint uncertainty of the static parameters and the latent path space estimates . In contrast, the AdPMCMC approach directly approximates these joint densities.

It is important to note that model parameterisation has important implications for the choice of static parameter priors in the population models used here. In this paper we have followed a common practice in the ecological literature, e.g., \citep{Clark2007}, wherein parameters such as the intrinsic rate of growth $r$ and carrying capacity $K$ are combined into new parameters $b_i$ such that the model on the log scale is linear in all (e.g., $M_{0,1,4}$) or some (e.g., $M_{2, 3}$) of the parameters. The $b_i$ for these models are often nonlinear functions of ecological parameters. We have observed that independent Gaussian priors on the $b_i$ induce strong prior dependence among the ecological parameters. The AdPMCMC approach does not require linearity in the parameters and can fit the underlying ecological parameters directly and efficiently. In other words our methodology is general and can be applied to any parameterization either linear or non-linear, under any chosen prior structure with or without dependence.

On the other hand, the more general $b_i$ formulation allows researchers to question how much support the data gives to parameter spaces consistent with the traditional ecological formulations. For example, Figure \ref{fig9:AnisusResults} shows that it is unlikely for both $b_0, b_2 > 0$. This corresponds to unchecked population growth in the latent model; this unlikely but not impossible result is precluded in the traditional formulation of the theta-logistic in terms of $r, K, \theta$ \citep{Morris2002}. The more general approach is useful because there are many ways to parameterize a population model's structure, and different ecological considerations may lead to quite different constraints. For instance, previous studies have proposed the following constraints in the theta-logistic model $M_2$: $b_3 > 0$ \citep{Ross2006, Ward2006}, $b_3 > -1$ \citep{Saether2008, Ross2009}, $\sgn{b_3} = \sgn{b_0}$ \citep{Polansky2009}. Figure \ref{fig9:AnisusResults} shows that these constraints may be unrealistic or unduly restrictive in the presence of both process error and observation noise. We argue that, wherever possible, it is better to estimate the probability of constraints rather than impose them, and the more general formulation provides a way to achieve this. Finally, we note that hyperpriors on the prior parameter values can also be considered in the AdPMCMC framework to address the usual concerns about prior sensitivity.

In our opinion the most important results of this study: a) are the confounding effects of observation and process errors on model selection;and, b) the ability of the AdPMCMC to mix efficiently over the complex ridges and multiple modes of the likelihood surfaces associated with at least some population dynamic models. In particular, it is clear that the development of adaption in the MCMC proposal used for the static parameters clearly improves mixing. Through the use of Bayes factors we have emphasised how important it is to consider a range of alternative models when seeking to understand the nonlinear and density dependent effects that drive population dynamics. It is important to recognise, however, that our limited understanding of these effects, reflected in the process error of a population model, coupled with our imperfect measuring devices, reflected in the variance of a observation model, may hamper our ability to distinguish between different plausible models. It is important that ecologists recognise this, particularly if a ``best fitting'' model to make predictions of population trajectories beyond the observations.

The AdPMCMC algorithm presented here is capable of efficient SSM inference in non-linear population dynamic models with complex likelihood surfaces. It thereby frees practioners from the potentially slowly mixing constraints of the SSM algorithms, particularly Metropolis Hastings within Gibbs, that are currently available. The AdPMCMC strategy can be readily generalised to other equivalent problems, and moreover, readily extends to include more complex SMC methods that also incorporate adaption and/or more realistic error variance structures. For these reasons, we believe the algorithm holds great promise in applied contexts.

\section{Acknowledgments}
We thank Prof. Arnaud Doucet for comments and suggestions during his research visit to CSIRO in Tasmania. The contributions of GRH and KRH were funded by the CMIS Capability Development Fund (CDF).GWP gratefully acknowledges the support of CDF during this research collaboration.

\bibliographystyle{plainnat}
\bibliography{working}

\begin{algorithm}[!ht]
\begin{algorithmic}[1]
\LINE \hspace{-1cm}\textit{Initialisation of SIR filter at iteration $(j+1)$ of the Markov chain}
\STATE{SIR particle filter for $N_{1:T}$: initialise $L$ particles $\left\{\left[n_1\right](j+1,i)\right\}_{i=1:L}$ via sampling from the priors in Section \ref{Priors}}

\FOR{$t=2,\ldots, T$}
\LINE \hspace{-1cm}\textit{Perform mutation of the $L$ particles at time $t-1$ to obtain new particles at $t$ via state evolution.}
\STATE{Sample the $i$-th particle $\left[n_t\right](j+1,i)$ from particle filter proposal according to state equation given in relvant model $M_i$ Section \ref{stateEqn}.
Example $M_3$ sample
\begin{equation}
\begin{split}
N_t \sim & N\left( 2\log{[N_{t-1}](j,i)} - \log\left([b_{4}](j+1,i) + [N_{t-1}](j,i)\right) + [b_0](j+1,i)\right) + \\  
& N\left([b_{1}](j+1,i)[N_{t-1}](j,i),[\sigma^2_{\epsilon}](j+1,i)\right) \\
\end{split}
\end{equation}
}

\LINE \hspace{-1cm}\textit{Incremental SIR importance sampling weight correction.}
\STATE{Evaluate the unnormalised importance sampling weights, $\left[\tilde{W}_t\right](j+1,i)$, for the $L$ particles, with the $i$-th weight given by
\begin{align}
\begin{split}
\left[\tilde{W}_t\right](j+1,i) &\propto  \left[W_{t-1}\right](j+1,i) \left[w_{t-1}\right](j+1,i)\\
&\propto  \left[W_{t-1}\right](j+1,i) p (y_{t}|\left[n_t,\bm{\theta}\right](j+1,i)),
\end{split}
\end{align}
}

\STATE{Normalise the importance sampling weights $\left[W_{t}\right](j+1,i)=
\frac{\left[\tilde{W}_{t}\right](j+1,i)}{\sum_{i=1}^{L}\left[\tilde{W}_{t}\right](j+1,i)}$}
\LINE \hspace{-1cm}\textit{Evaluate the importance estimate and resample adaptively.}

\STATE Calculate the Effective Sample size, $ESS =\frac{1}{\sum_{i=1}^{L}\left[W_{t}\right](j+1,i)^2} $

\STATE{If the Effective Sample size is less than 80\% then resample the particles at time $t$ using stratified resampling based on the empirical distribution constructed from the importance weights to obtain new particles with equal weight.}
\ENDFOR
\STATE Evaluate marginal likelihood $\widehat{p}\left(y_{1:T}|\left[\bm{\theta}\right](j+1)\right) = \prod_{t=1}^T
\left(\frac{1}{L}\sum_{i=1}^{L} \left[w_{t}\right](j+1,i)\right) $
\caption{Construction of optimal path space proposal for Model $M_i$ given by $\widehat{p}\left(n_{1:T}|y_{1:T},\left[\bm{\theta}\right](j+1)\right)$}

\label{SIR_filter}
\end{algorithmic}
\end{algorithm}

\begin{algorithm}[!ht]
\begin{algorithmic}[1]
\label{adaptice_MCMC}
\STATE Sample a realisation $u_1$ of random variable $U_1 \sim U[0,1]$
\LINE \hspace{-1cm}\textit{ Sample from the adaptive mixture proposal (\ref{AdaptiveProp}) }
\IF {$u_1 \geq w_1$}
\LINE \hspace{-1cm}\textit{ Sample $\left[\bm{\theta}\right](j+1)$ from the adaptive component of the mixture proposal of (\ref{AdaptiveProp})}
\STATE Estimate $\Sigma_j$, the empirical covariance of $\bm{\Theta}$, using samples $\{[\bm{\theta}](i)\}_{i=1:j}$.\;
\STATE Sample proposal $\left[\bm{\theta}\right](j+1) \sim N
\left(\bm{\theta};\left[\bm{\theta}\right](j) ,\frac{\left(2.38\right)^2}{d}\Sigma_j\right)$;
\ELSE
\LINE \hspace{-1cm}\textit{ Sample $\left[\bm{\theta}\right](j+1)$ from the non-adaptive component of the mixture proposal of (\ref{AdaptiveProp})}
\STATE Sample proposal $\left[\bm{\theta}\right](j+1) \sim N\left(\bm{\theta};\left[\bm{\theta}\right](j),
\frac{\left(0.1\right)^2}{d}I_{d,d}    \right)$
\ENDIF

\caption{Adaptive MCMC for static parameters $\bm{\Theta}$}
\end{algorithmic}
\end{algorithm}

\end{document}